# Network topology of the Argentine interbank money market

**Federico D. Forte**

*Economic Research Department, BBVA Bank Argentina*
E-mail: federico.forte@bbva.com

This version: September 2020

**Abstract**

This paper provides the first empirical network analysis of the Argentine interbank money market. Its main topological features are examined applying graph theory, focusing on the unsecured overnight loans settled from 2003 to 2017. The network, where banks are the nodes and the operations between them represent the links, exhibits low density, a higher reciprocity than comparable random graphs, short average distances and its clustering coefficient remains above that of a random network of equal size. Furthermore, the network is prominently disassortative. Its structural metrics experienced significant volatility, in correlation with the economic activity fluctuations and regulatory shifts. Signs of nodes' random-like behavior are detected during contractions. The degree distributions fit better to a Lognormal distribution than to a Poisson or a Power Law. Additionally, different node centrality measures are computed. It is found that a higher centrality enables a node to settle more convenient bilateral interest rates compared to the average market rate, identifying a statistical and economically significant effect by means of a regression analysis. These results constitute a relevant input for systemic risk assessment and provide solid empirical foundations for future theoretical modelling and shock simulations, especially in the context of underdeveloped financial systems.

**1. Introduction**

Financial entities exhibit a high degree of interdependence. They forge interlinkages via both sides of their balance sheets, which are essential for efficient financial intermediation. The 2008 global turmoil underscored the necessity of a more rigorous comprehension of the systemic risks associated with these interconnections among banks. Additionally, those events highlighted the central role played by interbank money markets for a correct functioning of financial systems and for the effectiveness of monetary policies.

Network analysis and graph theory provide an insightful framework to address these complex interrelationships. This approach allows us to better understand phenomena like contagion, network externalities, cascade failures, etc., which have been highly emphasized by recent literature specialized in financial stability [1, 2].

This paper provides the first comprehensive network analysis of the Argentine interbank market. Specifically, it examines the topological structure of the unsecured money market, known as *call market*. Banks conduct the management of a portion of their short-term liquidity positions through it. The average interest rate agreed, named *call rate,* represents a fundamental reference of the "cost of money" in Argentina. This short-term rate is a key benchmark for the determination of other longer-term rates in the economy. For this reason, the call market embodies one of the most direct transmission channels at disposal of the Central Bank of Argentina (BCRA) to implement its monetary policy. Hence the importance of examining the structural features of this network of loans, with the aim of assessing its stability and systemic risks.

Research on the Argentine interbank markets is extremely scarce. No previous work has focused on them, except for that of Anastasi, Elosegui and Sangiácomo [3], who studied the determinants of the call rate applying econometric methods for panel data. Therefore, the present study represents an initial milestone for the exploration of the domestic banking system from the vantage perspective of graph theory.

Thus, the call market is represented as a network, where financial institutions are nodes and the overnight loans among them are links. Its main structural features are analyzed from 1 January 2003 to 31 December 2017, based on a rich data set of daily transactions stored by the BCRA, in order to investigate if it shows similarities with stylized network models and to detect changes in their evolution over the years. This task enables us to draw conclusions, for instance, about its resilience to different types of disruptive events.

This line of empirical research has been growing extensively in the world since the early 2000s, along with the development of computational technologies and a larger availability of suitable data sets for applying these methods. Comparable studies were carried out on the money market networks of several countries, such as Italy [4, 5], U.S.A. [6], or Switzerland [7], just to name a few.

Argentina is a middle-income country with one of the smallest financial systems in South America. The country has experienced substantial macroeconomic volatility, coupled with sudden and sharp swings in financial regulation during the period under analysis (including financial repression and subsequent liberalization). Hence, the study of the Argentine financial network's dynamics in such a long and turbulent timeframe represents a good opportunity to gauge the responses of an underdeveloped system to multiple environments and distress events.

The 15-year time span addressed in this paper is one of the longest intervals examined thus far, when compared to the existing empirical studies on financial networks. We analytically subdivided this period into six different stages. This approach allows us to explore more accurately the variability experienced by the network structure throughout these years, which, to some extent, tended to move together with the macroeconomic fluctuations of Argentina's economy.

The next section explains the main institutional arrangements of the Argentine call market, the different monetary policy instruments and the minimum liquidity requirements imposed by the BCRA, which impact directly on money markets. Section 3 summarizes the most relevant findings of previous empirical literature on financial networks, providing a benchmark for our own estimates. Section 4 describes the database used. The methodological framework is explained in Section 5. The results are outlined in Section 6. Section 7 presents an econometric regression aimed at quantifying the effects derived from node centrality on banks' capability for negotiating more convenient interest rates on their bilateral transactions in the call market. Finally, Section 8 lists some concluding remarks and lines of research for future work.

## 2. The Argentine interbank money market

The call market is the Argentine traditional interbank market in which banks negotiate their liquidity positions with each other. The daily average interest rate of these transactions represents one of the most relevant short-term rates in the economy. The loans in this market are unsecured and are agreed between entities by telephone trading. Banks define bilaterally the interest rate of the transaction. Only institutions authorized by the BCRA can operate in this market. The vast majority of the loans are overnight, although a few longer-term transactions also take place. Financial institutions make a risk assessment of each possible counterparty and then define specific credit lines for each one (mainly, they determine the limit amount of money to be granted). Hence, when an entity needs liquid funds, it resorts in the first place to those banks with which it has credit lines available. This gives rise to repeated interactions between pairs of agents in the market.

The bilateral transactions are compensated through the real-time gross settlement system called "MEP" (stands for "*Medio Electrónico de Pagos*"), which is administered by the BCRA. The transfers of funds are not subjected to settlement risks because the monetary authority verifies, before the settlement of each loan, the existence of the required funds in the accounts involved.

In Argentina, there is another complementary market in which the financial institutions can negotiate their liquidity positions, known as "REPO market". In contrast, this is a secured market and transactions are conducted through an electronic platform. Nevertheless, it is mandatory to fulfill several costly conditions to operate in this market, referred to the volume of assets and equity of the bank (among others), which are often impossible to meet by a significant number of entities. These barriers to entry explain, at least partially, the substantial role played by the call market in the local financial system. Unfortunately, the necessary information to analyze the REPO market is not available yet, so this study will focus only on the call market.

Due to the call rate's key role as a benchmark for other interest rates, the BCRA has developed several instruments to influence its behavior. Since 2002 (in a context of a public debt default), the monetary authority started to issue its own short-term securities, called LEBAC and NOBAC. These securities were designed to absorb or provide liquidity from/to the market, affecting therefore the interest rates and monetary conditions of the economy. In addition, since 2004, the BCRA began to operate actively in the REPO market. A Central Bank's repo is a secured loan to a financial institution, while a reverse repo is the opposite transaction. Usually, these loans have a maturity of 1 or 7 days. Fig.1A shows the evolution of the call rate during these years, jointly with the interest rates of the BCRA's most relevant monetary policy instruments.

## 2.1. *Macroeconomic context*

Money markets in Argentina have faced wide fluctuations during the years under examination, along with the macroeconomic volatility experienced by the country. For that reason, the time period studied is analytically subdivided into six stages, defined according to the development of exogenous factors that affected crucially the interbank market (Fig.1B). This approach is useful to detect similarities, breaks and continuities in the evolution of its structural features over time.

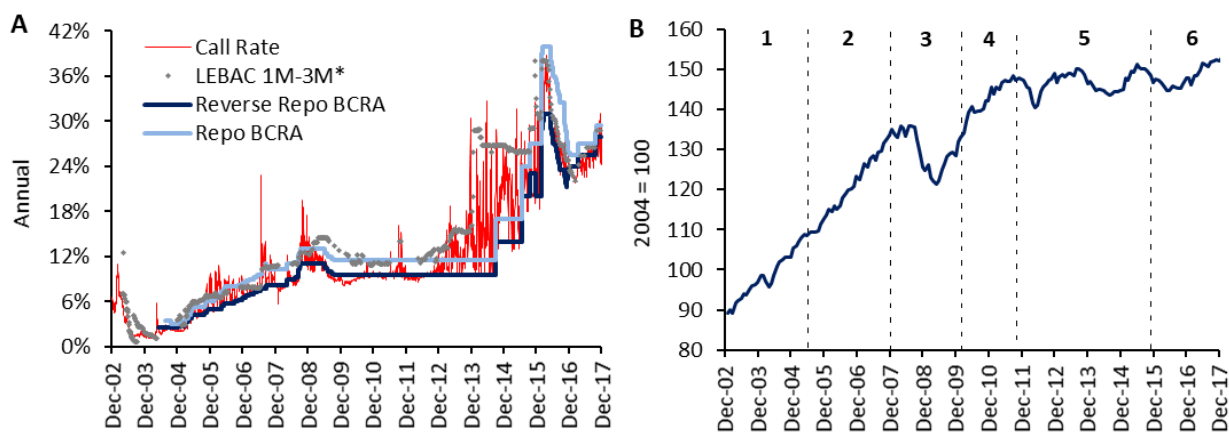

Fig.1. (**A**) Call rate and interest rates of monetary policy instruments (source: BCRA). *The Figure includes only the interest rates of the LEBACs with the shortest duration in each moment, provided that they are shorter than 105 days. (**B**) Argentina Monthly Economic Activity Estimator (EMAE), seasonally adjusted (source: INDEC). Vertical spotted lines demarcate the six different time stages considered for the analysis.

The first stage comprises the lapse before the public debt restructuring, between January 2003 and June 2005. The economy was starting to recover from the deep 2001-02 crisis and the financial system was facing many obstacles to its functioning, because of the widespread bank runs suffered in the previous years and due to the debt default, which hindered local operations of international entities. Within this first stage, the year 2003 is notably different from 2004, for two main reasons. In general, the economic situation of Argentina was considerably healthier during the latter, but the other relevant factor was that in 2004 the REPO market was established, which provided additional tools for managing liquidity to a still weakened domestic financial system.

Stage 2 is defined between July 2005 and December 2007, a period in which the economy was buoyant, after the debt restructuring and the surge in commodity prices. Then, Stage 3 includes the outbreak of the 2008 global crisis and its subsequent impact on Argentina (it is delimited between January 2008 and February 2010). The main effects of the crisis took place in the first half of 2009, and it was not until the second half of that year that the economy began to rebound. The fourth stage reviews the recovery from the crisis, characterized by a strengthened economic activity together with the worsening of both fiscal and external deficits ("twin deficits"). It is defined between March 2010 and October 2011.

By the end of October 2011, the government established harsh FX controls, introducing radical changes in regulatory frameworks, especially in the financial system. Capital mobility was strongly restricted and many regulations were imposed on banks' interest rates. Gross Domestic Product (GDP) has stagnated since then, giving rise to a period of recurrent macroeconomic recessions. The fifth stage covers these years, between November 2011 and November 2015. The call rate's volatility exacerbated during that lapse (see Fig.1A). Since December 2015 until the end of the sample period, FX and interest rate controls were completely relaxed. These last two years are included in Stage 6. An Inflation Targeting regime was established, and the BCRA used a monetary policy rate as its main instrument to manage the monetary conditions of the economy. In this context, the call rate reduced dramatically its volatility and progressively resumed a more similar behavior to the one displayed before 2012.

Table 1. *Analytical time stages*

| Stage | Period | Main Events |
|---|---|---|
| 1 | January 2003 - June 2005 | Debt default. Beginnings of economic recovery. |
| 2 | July 2005 - December 2007 | Debt Restructuring. Bouyant economy. |
| 3 | January 2008 - February 2010 | Global financial crisis. |
| 4 | March 2010 - October 2011 | Macroeconomic recovery. Twin deficits widening. |
| 5 | November 2011 - November 2015 | FX-market restrictions. Interest rate controls. |
| 6 | December 2015 - December 2017 | FX-market and financial liberalization. Inflation Targeting Regime. |

The number of financial entities decreased almost monotonically during the years under analysis. Starting from a total of 100 in January 2003, only 77 were active in December 2017 (Fig. 2). We employ a usual classification in the domestic financial system to study the dynamics of different types of banks in the network. It divides them into four subgroups, according to the owner of the institution: State-Owned Banks (SOBs), Domestic Private Banks (DPBs), Subsidiaries of Foreign Banks (SFBs) and Non-Bank Financial Institutions (NBFIs). This classification, based on the structure and ownership of the institutions' equity, is also useful as a proxy for the specific type of financial businesses run by each entity. The number of institutions decreased in all the subgroups, but the SFBs declined the most, going from 28 banks in 2003 to 16 in 2017.

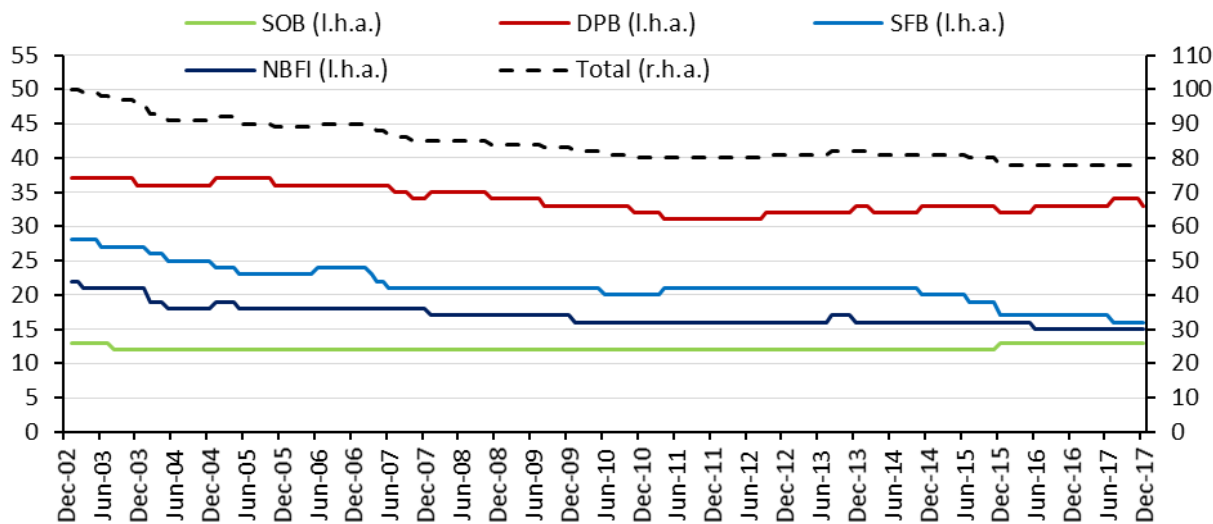

Fig. 2. Number of entities in the financial system, by type (source: BCRA). DPBs: Domestic Private Banks; NBFIs: Non-Bank Financial Institutions; SFBs: Subsidiaries of Foreign Banks; SOBs: State-Owned Banks.

2.2. *Banks' minimum reserves*

Minimum liquidity requirements imposed by the BCRA affect crucially the interbank money market, both directly and indirectly. In Argentina, they are defined according to the deposits of each entity, with different requirements depending on their maturity.

Liquidity reserves are computed as the monthly average of banks' current account deposits at the BCRA. From 2003 to 2017, these requirements averaged 14% of total deposits, varying between 11% and 17% (considering all kinds of domestic currency deposits). At the end of each month, the monthly average of banks' current account balances at the BCRA must exceed these minimum levels to avoid financial penalties. Additionally, since 1997, a mandatory daily liquidity minimum has also been established, defined as a proportion of the monthly requirement (most of the time it was 50%). These regulations impact substantially on the interbank markets, since financial entities tend precisely to resort to these markets to adjust their liquidity excesses or shortages, always considering the constraints imposed by the monetary authority.

## 3. Network theory and empirical interbank markets

Over the last decade, academic researchers as well as policy-makers have devoted a good deal of attention to issues associated with financial system's interconnectedness and its implications in terms of stability and vulnerability [8]. There is no clear consensus on whether a completely interconnected financial network reduces contagion and domino effects [9, 10] or fuels them [11]. Interconnectedness among banks improves risk diversification, but at the same time it makes them more prone to contagion. In any case, from a financial stability perspective, there is broad agreement on the idea that banks should neither be *too-big-to-fail* nor *too-interconnected-to-fail* [12], as these types of institutions entail a potential source of vulnerability for the whole system.

Boss *et al* [13], for Austria, and Inaoka *et al* [14], for Japan, carried out the first topological analyses of real interbank networks comparable to the one proposed in this paper. Thereafter, several pieces of research have been developed in this line. The Appendix summarizes the main results of 27 empirical studies on financial networks of 18 countries (including the topological measures computed here for Argentina), in order to have benchmark metrics for our findings.

Overall, real-world interbank networks are sparse, which means that they are far from complete. In almost all empirical cases the average reciprocity is higher than the density of the network, so the connections tend to be more reciprocal than in random graphs [15]. This fact denotes that financial institutions prefer to interact with those agents with whom they have already established relationships in the past. Hence the relevance of taking into consideration the presence of stable interconnections when assessing and modelling the behavior of financial entities.

Furthermore, interbank networks exhibit higher clustering coefficients than random graphs of equal size, but substantially lower than regular networks. They also display, in general, short distances (between 1.5 and 4, on average). Absolutely all the pieces of research reviewed conclude that interbank networks show disassortative mixing, which means that nodes with many edges tend to be connected to nodes with relatively less links, and vice versa.

There is consensus that financial networks follow heavy-tailed distributions. Many studies ascertain that degree distributions fit reasonably well to a Power Law (with an exponent between 2 and 3.5), which means that these networks can be described as scale-free [16]. Consequently, it is usual to find few banks with an extraordinarily high number of edges, coupled with a myriad of nodes far less interconnected. The number of agents in financial networks tends to be relatively low (much smaller, for example, than in biological or social networks), a fact that hampers significantly the statistical analysis of their degree distributions.

Depending on the specific type of data analyzed, three subgroups of financial networks can be clearly identified across literature: 1) balance sheet exposures; 2) payments; and 3) transactions in the interbank money markets. This paper is focused on the latter. This kind of networks are usually smaller than payment networks and are, on average, the sparsest among the three subgroups. In addition, they exhibit significantly lower clustering coefficients than payment systems, given that in general their values do not exceed 0.2, while in the second case the average clustering is around 0.5. The networks based on balance sheet exposures also show clustering coefficients slightly smaller than payment systems.

## 4. Data

The BCRA stores daily data about all the transactions carried out in the call market. The information available includes the lender and the borrower entities, the amount of money, the loan maturity, the currency involved, the agreed interest rate and the type of rate (fixed or variable). Our sample consists of 314,188 loan operations, conducted between 2 January 2003 and 29 December 2017, by 99 different entities (12 SOBs, 41 DPBs, 27 SFBs and 19 NBFIs). 99.3% of the transactions were settled in Argentine pesos and 88.8% were overnight. With the aim of getting comparable results to those of other empirical studies on financial networks over the world, the sample is restricted to consider only the overnight loans in pesos with fixed interest rate. This subset comprises 278,497 operations (88.6% of the entire dataset).

For the construction of weighted networks, all the weights involved are based on amounts of money expressed in millions of constant pesos of 2017, that is, in real terms (constant purchasing power)[1]. The data on total deposits, assets and liquidity of financial institutions (used in the regressions of Section 7) emerge from information collected by the BCRA.

## 5. Notation and methods

Since the minimum liquidity requirement established by the BCRA for banks is based on the average amount of reserves over a whole month, the monthly networks appear to be a better approximation than the daily networks to grasp the genuine lattice of relationships among banks emerging from their liquidity management. This approach can also be understood noting that all the transactions in the call market are based on the previous existence of open credit lines among banks, which set up a "latent" network of interrelationships. Every day, some links "activate" and others do not, but they remain available in case of need.

---

[1] The Consumer Price Index (CPI) was obtained from different sources: a) January 2003 to November 2006 and May 2016 to December 2017: INDEC; b) December 2006 to April 2011: average of provinces' CPIs; c) May 2011 to July 2012: National Congress; d) August 2012 to April 2016: City of Buenos Aires' CPI.

5.1. *Topological indicators: size and connectivity*

To analyze the call market as a network, each node represents a bank that carried out at least one transaction in the month, while edges are created when at least one operation was settled between a pair of entities during such lapse. A directed network is then defined by this set of nodes (N) and edges (M). The components ($a_{ij}$) of its Adjacency Matrix (a $N \times N$ matrix denoted by A) are 1 if $|w_{ij}| > 0$ and 0 otherwise, where $w_{ij}$ is the average amount of money loaned to the bank $j$ by the bank $i$ in the month. Therefore, a link is incoming to the borrower and outgoing from the lender.

The distance ($d_{ij}$) between two nodes $i$ and $j$ is defined as the number of edges along the shortest path connecting them. The longest distance in the network is called "diameter". The density or "degree of completeness" ($\delta$) of a network quantifies the percentage of the potential links that actually exist, given the number of nodes in the graph. It is computed as follows: $\delta = (\sum_{ij} a_{ij})/(N(N-1))$.

It is often relevant to know if edges are reciprocal, i.e., to find out to what extent the links that go from node $i$ to node $j$ are also directed in the opposite way. The standard measure of "reciprocity" (R) is the following: $R = \sum_{ij} a_{ij} a_{ji} / M$. However, this indicator does not take into account that denser networks tend to have a higher number of reciprocal links, due exclusively to random reasons [17]. To address this issue, $R$ can be adjusted for the density of the network: $R_{adj} = (R - \delta)/(1 - \delta)$. Values of $R_{adj}$ above (below) zero imply a larger (lower) reciprocity than a random network with the same density.

The "degree" of a node $i$ ($k_i$) reflects the number of nodes with which it is connected. In the case of directed graphs, the in-degree ($k_i^{in}$) is the number of nodes with which node $i$ has incoming edges, while the out-degree ($k_i^{out}$) quantifies the node $i$'s outgoing links. Similarly, the *node strength* ($s_i$) is the sum of the weights ($w_{ij}$ and $w_{ji}$) of all the edges of a node; that is, the sum of the money involved in all the links of a given vertex $i$. It is convenient to assess in a different way the relevance of the entities that operate large liquidity flows per month, with respect to the entities that may be connected to many others (i.e., display a high degree) but through low-value operations. The nodes' in-strength ($s_i^{in}$) and out-strength ($s_i^{out}$) will be analogously computed, but weighting the edges only by the funds borrowed or loaned, respectively.

The *assortativity* of the nodes ($\rho_{kj}$) reflects their preference between the option of being connected with others of a similar degree to themselves or relating to a greater extent with those that exhibit a different degree. Many ways to compute this metric were developed, but we use the Pearson correlation coefficient between the degrees of nodes that share links, in line with Newman [18]. Consequently, $\rho_{kj}$ ranges from -1 to 1. If it is positive, the network is said to show an assortative behavior (also called "homophily"), meaning that the nodes tend to be connected to others of a similar degree. If it is negative, the network is "disassortative", implying that low-degree nodes tend to be attached to high-degree nodes and vice versa. Disassortative networks are particularly vulnerable to targeted attacks on their highest-degree vertices, while assortative networks proved to be more resilient to them [18].

The *clustering coefficient* of node $i$ is defined as follows: $c_i = \frac{1}{k_i(k_i-1)/2} \sum_{j,h} a_{ij} a_{ih} a_{jh}$. It is a measure of the probability that two nodes which are neighbors of a same node also share a link themselves. A high clustering reveals the existence of stable relationships among the nodes, with all the potential consequences that it entails, which can be either positive (e.g., an enhanced resilience to random shocks) or negative (e.g., higher level of contagion when other banks fail).

## 5.2. *Centrality and concentration*

Centrality measures are key to detect *too-interconnected-to-fail* vertices, and consequently to estimate the potential vulnerability of a network. Several metrics have been developed, based on different approaches. Degree Centrality ($k_i$) is one of the most basic ones. According to it, a node is more relevant if it has a higher degree. If a bank is connected to many others, its failure could impact on them directly. For some specific purposes, it is important to examine the nodes' out-degree ($k_i^{out}$) and in-degree centrality ($k_i^{in}$). Similarly, the Strength Centrality can be analogously interpreted, either considering the banks' total strength ($s_i$), in-strength ($s_i^{in}$) or out-strength ($s_i^{out}$).

The Closeness Centrality (CC) of a node is based on how many intermediaries are required to pass through in order to reach it. This measure is associated with the capability of a node to spread contagion, and it is calculated as follows: $CC(i) = (N-1)/(\sum_j d_{ij})$. In contrast, Betweenness Centrality is related to the strategic location of a node on the network's communications paths. In the case of the call market, this type of centrality reflects the influence of a node on the liquidity channels within the system. Betweenness Centrality reveals how fast potential shocks can spread through the network, while other measures, like degree or closeness centrality, account for the probability of amplification of shocks to the neighbors of each vertex [19]. The Betweenness Centrality (B) of a node $i$ is defined as: $B(i) = \sum_{i \neq j \neq h \in N}[(\sigma_{jh}(i))/\sigma_{jh}]$, where $\sigma_{jh}$ is the number of shortest paths between $j$ and $h$, and $\sigma_{jh}(i)$ is the number of shortest paths between $j$ and $h$ that pass through node $i$. Dividing $B(i)$ by (N-1)*(N-2), the measure is normalized, so that it can be applied homogeneously to graphs of different sizes.

The last centrality measure examined is the Eigenvector Centrality, proposed by Bonacich [20]. As its name suggests, the centrality value of node $i$ is given by the $i$th entry of the eigenvector ($e$) associated to the largest eigenvalue ($\lambda$) of the graph's adjacency matrix (A). This measure takes into consideration the centrality of a node's neighbors to compute its own centrality. It can be understood as the weighted sum of the direct and indirect connections of the node, at any length.

All these measures are defined in such a way that a higher value is always interpreted as a larger centrality of a node in the network. Based on these metrics, Section 7 assesses the effect of a higher centrality on the banks' capability to agree more convenient interest rates in the call market. To this purpose, an econometric regression is estimated by Ordinary Least Squares (OLS), including control variables to remove potential sources of endogeneity (in line with the procedure applied by [6]). The robustness of the resulting coefficients is tested by examining alternative specifications (computing heteroskedasticity-robust standard errors). All the daily transactions are considered individually for these exercises (not their monthly averages). Comparable studies were carried out on the Fed Funds Market in U.S.A. [6], and the interbank markets of Norway [21], Switzerland [22] and Germany [23]. All these contributions verified the existence of a positive effect derived from a higher centrality on the ability to agree better interest rates in money markets. However, it is worth noting that all those studies are based on developed financial systems.

The centrality analysis is complemented by an assessment of the concentration of liquidity flows. The typical Herfindahl-Hirschman Indices are computed to measure the concentration in the lenders' side of the market -HHI(L)- and among the borrowers -HHI(B)-:

$$HHI(L) = \sum_i \left(\frac{v_{ij}}{V}\right)^2 \; ; \; HHI(B) = \sum_i \left(\frac{v_{ji}}{V}\right)^2$$

$v_{ij}$ and $v_{ji}$ denote, respectively, the total amount of money loaned and borrowed by each entity $i$ in a specific month (expressed in real terms). $V$ refers to the total traded volume in the network ($V = \sum_i v_{ij} = \sum_i v_{ji}$). The HHI is often used to measure market power. It ranges from 0 to 1. A higher value indicates a greater concentration of liquidity among few participants. To obtain comparable measures across time and different network sizes, this index is normalized by adjusting it according to the changes in N: $HHI^* = ((HHI - 1/N))/(1 - 1/N)$.

## 5.3. Degree distribution

In random graphs, most of the nodes have a similar degree, near to the average. But real networks hardly ever show this property [16]. In many contexts their degree distributions follow a Power Law, in which case they are called *scale-free networks*. Mathematically, a quantity $k$ obeys a Power Law if it is drawn from a Probability Density Function (PDF) as follows: $P(k) \propto k^{-\alpha}$, where $\alpha$ is a constant parameter. One remarkable feature of this distribution is that it frequently originates extreme values very far from the mean, exhibiting heavier tails than Poisson, normal or exponential distributions. This gives rise to networks in which few highly interconnected nodes coexist with many low-connected nodes.

Scale-free networks are *robust-yet-fragile* structures [24]. They exhibit a surprising degree of tolerance against random removals of a relatively large number of nodes (*robustness*). However, this error tolerance is coupled with a high susceptibility to targeted attacks. These networks break rapidly into isolated fragments when a few of the most connected nodes fail (*vulnerability*). This attribute has critical implications in terms of systemic fragility of interbank networks. If a financial network displays this behavior, then a rigorous identification of the main agents becomes a priority task for central banks and regulators. Instead, random graphs show the converse risk structure. They easily absorb targeted attacks but tend to fall apart swiftly when random failures occur.

Thus, one objective of this paper is to verify if the empirical degree distributions of the monthly networks behave like random graphs (which are best described by a Poisson), or if the empirical data fits better to a "heavy-tailed" distribution, such as, for example, a Power Law or Lognormal. In the latter case, "unusual" or disruptive events exhibit a relatively higher probability than in the random contexts, a fact with significant implications in terms of systemic risk.

The methodology proposed by Clauset, Shalizi and Newman [25] is applied to this purpose, as it has proven to achieve more robust results than the other existing techniques. It is often the case that degree distributions follow a Power Law only in the tail, i.e., for values above a certain lower bound ($x_{min}$). Hence, an unbiased selection of $x_{min}$ constitutes a crucial prerequisite to estimate the other parameters of the fitted distributions, which are in turn estimated by the Maximum Likelihood (ML) method. Clauset *et al* [25] suggest choosing the value of $x_{min}$ that minimizes the differences between the distribution of the empirical data and the best-fit theoretical model (regardless of the model selected). For non-normal data, the most usual method to measure the distance between two probability distributions is the Kolmogórov-Smirnov (KS) statistic:

$$KS = \max_{x \geq x_{min}} |D(x) - P(x)|,$$

where $D(x)$ is the Cumulative Distribution Function (CDF) of the data, for values above $x_{min}$, and $P(x)$ is the CDF of the hypothetical model fitted by ML, for the same region $x \geq x_{min}$. The proposed estimate of $x_{min}$ is then the value that minimizes KS. The variances of the parameter estimates (both for $x_{min}$ and the others) are computed by the non-parametric bootstrap method.

However, once the parameters of a specific distribution are estimated to fit the data, it is also necessary to analyze whether that distribution represents a plausible description of the data or not. To this purpose, Clauset *et al* [25] suggest a goodness-of-fit test which generates a p-value that measures the plausibility of the hypothetical model. In this test, the "distance" (the KS statistic) between the distribution of the empirical data and the hypothesized model is compared with distance measurements for, say, 3,000 synthetic datasets generated from the same proposed theoretical model (i.e., 3,000 synthetic KS statistics). A p-value is then defined as the fraction of the synthetic distances that are larger than our original KS. If this p-value is sufficiently large, the hypothesized distribution would represent a plausible fit to the data. The authors choose a p-value ≤ 0.1 as threshold for rejection, but they also suggest a less stringent rule of p-value ≤ 0.05.

Finally, the performance of the different proposed models is then assessed by comparing the log-likelihood levels associated to each fitted distribution, based on the empirical data and the parameter estimates. For instance, if the log-likelihood resulting from the Lognormal fit is higher than the one derived from the other proposed distributions, that would indicate that the Lognormal describes better the data set (even in the possible case that none of the hypothesized distributions was rejected by the goodness-of-fit test described in the previous paragraph).

A note of caution is in order. Scale-free properties tend to emerge in large graphs, but financial networks are usually relatively small (compared to social or biological ones, for example). This fact can bias the process of fitting a set of observations to a theoretical distribution [25]. For this reason, many recent empirical studies focused on the less ambitious task of detecting "heavy-tailed behaviors", testing fits with several fat-tailed distributions and not only Power Laws [26].

**6. Network analysis of the overnight money market**

6.1. *Size and connectivity*

The Argentine interbank network is relatively small (Fig. 3). However, this characteristic is shared by overnight loans networks in general, as they tend to be tinier than graphs based on balance sheet exposures or payments. If we circumscribe only to loans networks, the Argentine is not the smallest (cf. the Appendix), but it is substantially tinier than other developed markets, for instance, the Fed Funds network in the U.S.A. [6] or the Italian Overnight Money Market [26].

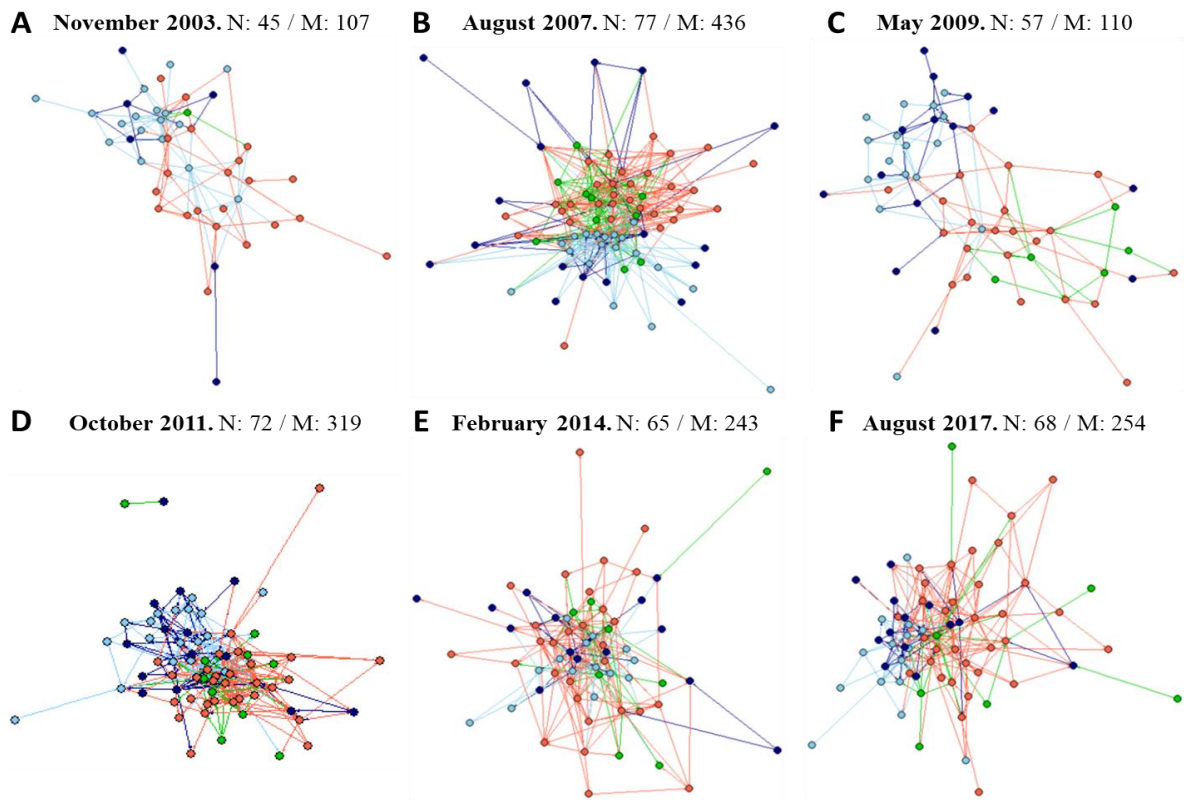

Fig. 3. Argentine interbank networks (monthly average). Each node represents a financial entity (green: SOBs; red: DPBs; light blue: SFBs; dark blue: NBFIs). Each edge denotes the existence of at least one loan settled between a pair of entities during the month, and its color is defined by the lender entity. The visualization layout was computed by the Fruchterman-Reingold algorithm.

From 2003 to 2017, the monthly networks had an average of 65±6 nodes and 237.5±73 edges (considering the directed graphs). These two variables underwent large fluctuations, mainly between 2003 and 2011 (Fig. 4A). Stage 1 is the most volatile period, when the financial system was still recovering from the 2001 crisis and the debt default was not solved yet. The interbank network started to grow considerably in 2004. The percentage of active banks in the call market with respect to the total number of entities in the system increased from 55% in 2003 to an average of 84.5% in 2007, and then stabilized at around 81.5% during the rest of the period (Fig. 4B).

The number of nodes and links reached peaks in 2007, during Stage 2. When the 2008 global crisis broke out, the networks suffered a sizable contraction. The graphs remained shrunk until the beginnings of 2010. Then, they started to grow again until 2011, when they got to the levels at which they would then stabilize until 2017, at around 66 nodes and 243 edges. Since 2011, the network showed substantial stability (in line with the country's economic stagnation during the same period), only paused by a transitory decline in early 2012, in a recessive context, after the introduction of severe FX controls that hindered the normal functioning of the financial system.

Nodes and edges tended to move together. In fact, from a log-log regression it is inferred that $M \propto N^{3.45}$, which implies that the average degree of the network -$\langle k \rangle$- increases with order $N^{2.45}$ (Fig. 4C)[2]. This result contrasts with theoretical papers that assume that $\langle k \rangle$ remains fixed over time [27]. This log-log relationship was found in almost every time stage, with the peculiarity that during the 2008 crisis the intercept seems to lie in a lower value than the average.

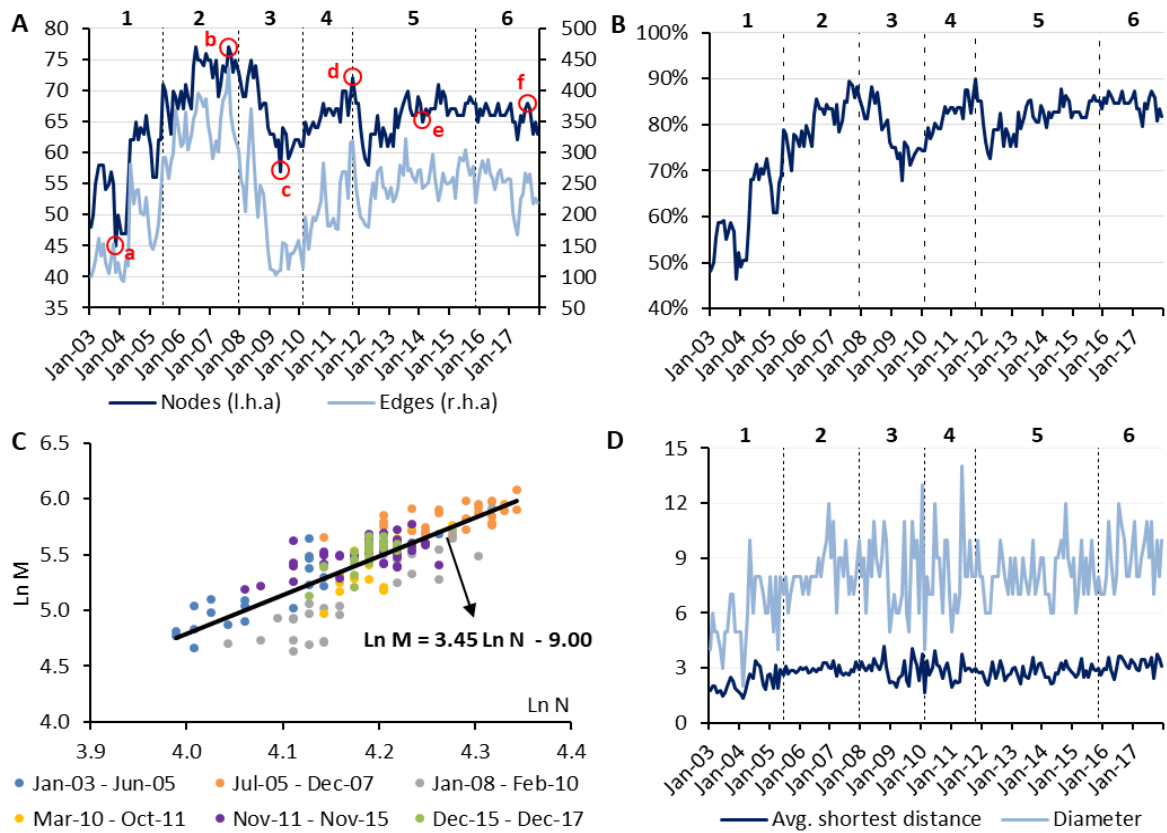

Fig. 4. **(A)** Active nodes and edges in the call market (monthly average). The red circles signal the networks shown in Fig. 3. A specific month of each stage was selected in order to display an introductory visual approximation of the network's structural changes throughout the period under analysis. **(B)** Active nodes in the call market (as percentage of total entities in the financial system). **(C)** Log-Log relationship between N and M. The horizontal axis depicts the logarithm of the number of nodes in each monthly network. The vertical axis shows the logarithm of the number of edges. **(D)** Average distance and diameter of monthly networks.

The average distance was 2.8±0.5, while the diameter oscillated around 7.9±1.9. These two metrics had proportional trajectories and roughly co-moved with the number of active nodes and edges (Fig. 4D). The average distance generally remained stable between 2 and 3, with few exceptions, in line with the typical range displayed by financial networks over the world (c.f. the Appendix). The diameter followed a similar path, though it was substantially more volatile.

---

[2] This M-N elasticity is significantly higher than the one found in the Italian overnight market ($M \propto N^{1.5}$) [26].

Overnight loans networks usually exhibit low density [12] in comparison with networks studied in other sciences (e.g., biology, physics, social networks). The Argentine interbank market is not the exception, since only 5.5%±1.1% of the potential edges actually exists. The monthly networks' density behaved similarly to the other topological measures already described (Fig. 5A). The reciprocity coefficient was on average 7.9%±3% and exceeded the level of density in 77.2% of the months. These values are in line with other real-world interbank networks, although the Argentine figures are among the lowest. The reciprocity experienced significant volatility, hitting a maximum of 17.7% in September 2011 and a minimum of 1.8% in early 2010. $R_{adj}$ stood at 2.5%±3.1%, implying that the network tends to show higher reciprocity than a random graph of equal degree of completeness. This means that banks are prone to establish two-way links between each other for non-random reasons, highlighting the importance of accounting for the existence of steady relationships within the call market (at least when evaluating its systemic stability).

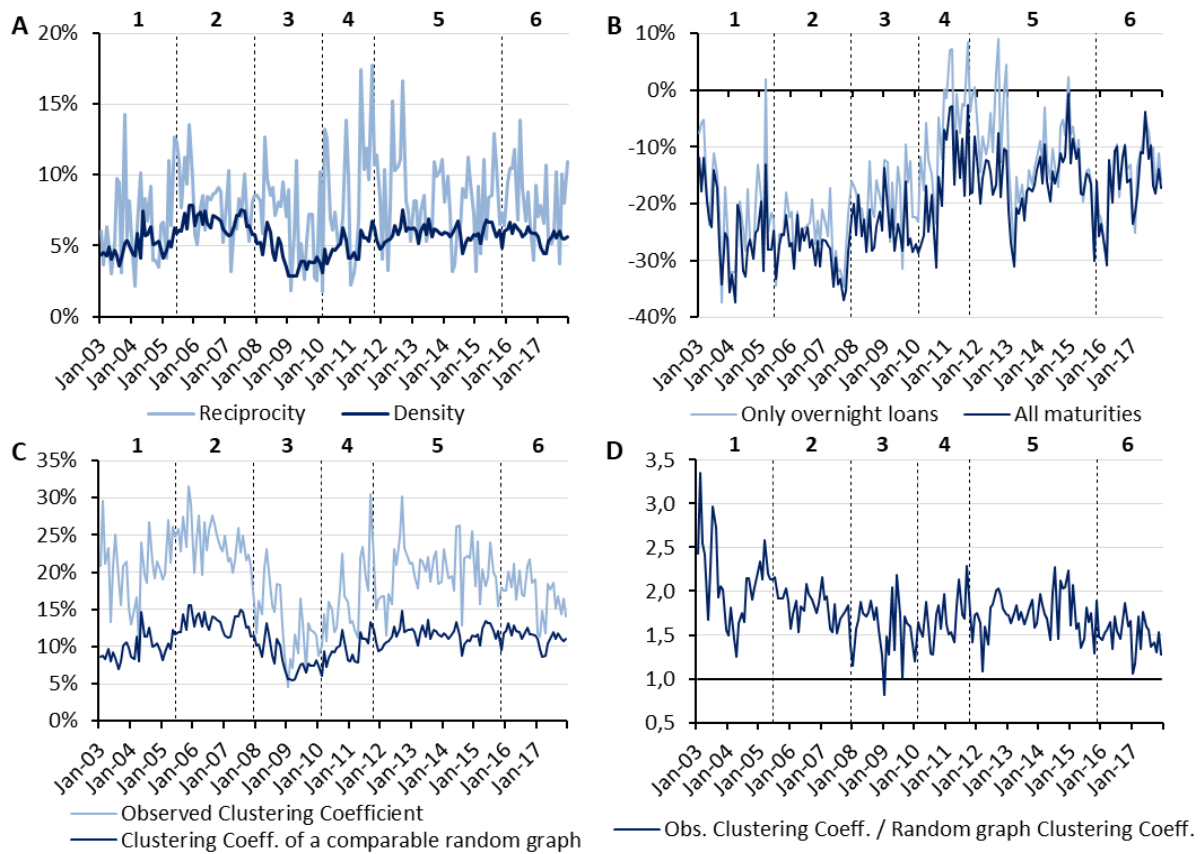

Fig. 5. **(A)** Density and reciprocity of monthly networks. **(B)** Assortativity coefficient, according to the maturity of the loans considered. **(C)** Clustering coefficient. The observed average clustering coefficient in the empirical networks is compared with the clustering coefficient that would emerge from a random graph with the same number of nodes (N) and with the same average degree ⟨k⟩. The average clustering coefficient of a random network is equal to: $\langle k \rangle / N$. **(D)** Relative clustering coefficient.

The assortativity coefficient averaged -16.3%±9.4% (Fig. 5B), which means that the network is disassortative. Low-degree banks are more likely to interact with high-degree banks than with other low-degree ones. It is the typical behavior found in all the other financial networks over the world. This disassortative mixing intensified during the first booming years (2003-07). The assortativity coefficients were negative in the 94% of monthly networks. Only in 10 months (out of the 180 analyzed) the market displayed a sporadic assortative behavior, which arose mainly when the reciprocity hit maximums, in 2011-12. However, when all the loan maturities are considered (i.e., not only overnight), networks remain always disassortative. This fact reinforces the idea that high-degree entities tend to create more links with low-degree ones, but it additionally suggests that the transactions settled with low-degree banks frequently have longer maturities.

The call market exhibited clustering coefficients systematically above those of a comparable random graph throughout the period, with only one exception in January 2009 (Figs. 5C and 5D). The creation of relatively stable (non-random) clusters is a usual practice in financial networks, due to the propensity to establish lasting relationships among financial agents. This conduct allows them to tackle problems related to moral hazard and adverse selection. Nevertheless, at the same time the need of risk diversification puts an upper limit to clustering levels. This trade-off explains why clustering coefficient in interbank networks tends not to reach comparatively high values in relation with other real-world networks (e.g., biological or linguistic networks). In broad terms, the Argentine case follows these guidelines, with an average clustering of 19%±5.1%.

In summary, the call market's main topological measures experienced a significant volatility, mostly between 2003 and 2010. Starting from very low levels in 2003, the number of nodes and edges, the diameter, average distance, density and reciprocity of the network grew altogether in simultaneous with the GDP until 2007. During this expansion, disassortativity increased, as also did the clustering coefficient. The global crisis triggered a strong contraction in all these metrics, and banks' reciprocity and clustering tended towards those of a random network. In 2010-11 the indicators rebounded sharply, with the peculiarity that reciprocity hit historical peaks along with transitorily positive assortativity coefficients. This singular behavior distinguishes that recovery phase from the previous growth period of 2003-07. In 2011, the topological metrics reached levels that would then remain stable on average until 2015. Not many significant changes were witnessed in 2016-17, except for an incipient decrease in density and relative clustering coefficient.

6.2. *Centrality and concentration*

Financial entities exhibited an average total degree of 7.1±1.8, which fluctuated proportionally to the number of edges. This indicator was clearly heterogeneous among the different types of banks. SFBs were the group with the largest average degree since 2003 until the global crisis, after which DPBs took over this central role in the market until 2017. Meanwhile, NBFIs were the group with the lowest average degree during all the period, playing a more peripheral role.

The different types of entities assumed contrasting positions in the network. SOBs had an average out-degree always higher than their in-degree, becoming the prime liquidity providers of the market. These banks held the highest average out-degree of the network between 2004 and 2008, the most buoyant years in terms of the market's activity. DPBs were important liquidity providers too, essentially between 2012 and 2017. In the opposite side, SFBs were the main borrowers, displaying the largest average in-degree almost throughout the period. NBFIs showed, in general, a higher average out-degree than in-degree, except in the last stage (Figs. 6A and 6B).

From the perspective of the node *strength* (Figs. 6C and 6D), conclusions do not differ significantly, but some peculiarities deserve to be highlighted. During the most dynamic moments of the network, between 2005 and 2007, the main players of the market were fundamentally the SOBs, as lenders, and the SFBs, as borrowers. DPBs also performed a key role in both sides of the market (but with less predominance), while NBFIs displayed a substantially higher out-strength than in-strength. With the outbreak of the global crisis, the average node strength of the system collapsed, from ARS 917±140 million to nearly 260±43 million in 2009. SOBs were the group of banks which suffered the sharpest decline. Since then, no group consolidated as the most central from the out-strength perspective. Even though since 2010 the average node strength recovered to ARS 370±60 million and remained around those levels until 2017, no group emerged as the most central liquidity provider in the market. In terms of in-strength, SFBs stood as the main borrowers almost every month. They also experienced a decay during the crisis, but nonetheless they bounced back relatively fast and kept their hegemonic role in the following years.

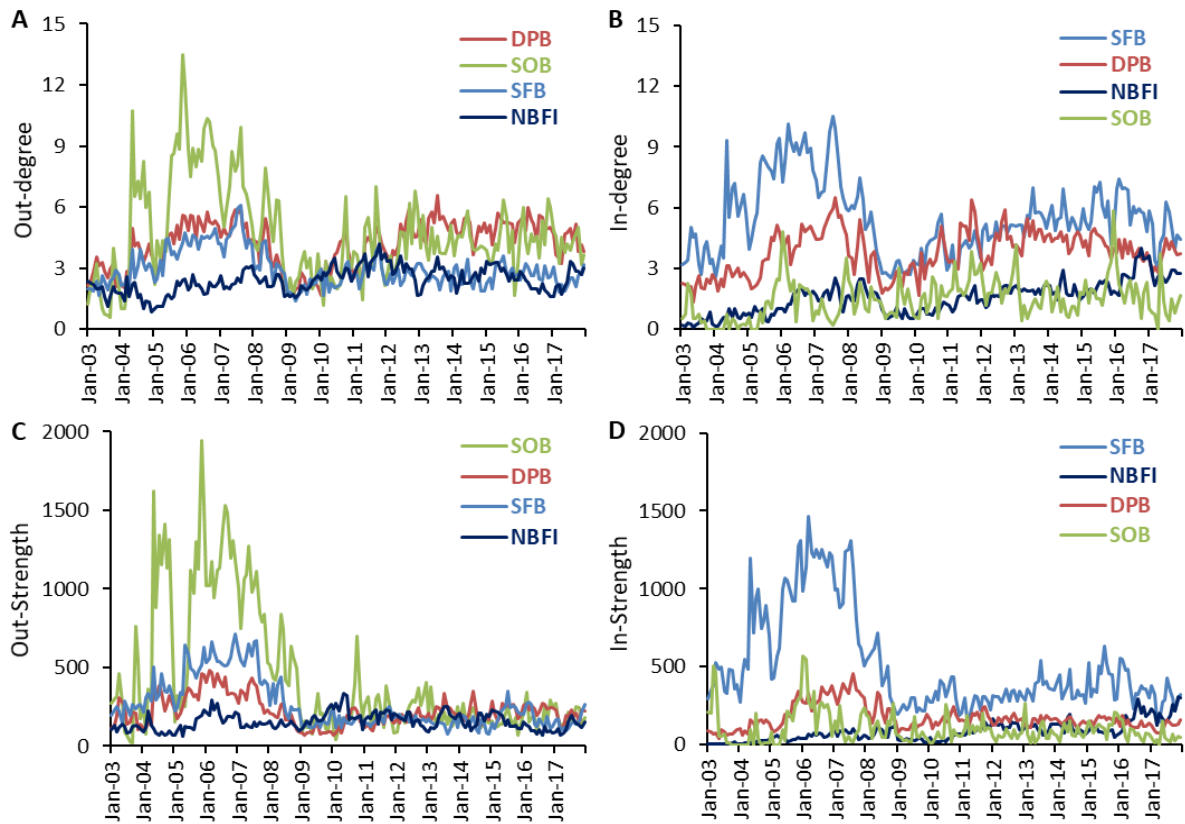

Fig. 6. **(A)** Average out-degree. **(B)** Average in-degree. **(C)** Average out-strength. **(D)** Average in-strength. The four figures show the monthly average values for each subgroup of entities. Node strength is expressed in ARS million (constant prices of 2017).

The network was more concentrated between 2003 and 2009 than in the succeeding years, more markedly in the case of borrowers -HHI*(B)- than lenders -HHI*(L)- (Fig. 7A). This result is consistent with the node strength's evolution. Both HHIs fell (i.e., the market reduced its concentration levels) over the years. Starting from about 10% on average in 2003-04, after the global crisis those indices plateaued until 2017, around 4.6% for lenders and 4.8% for borrowers.

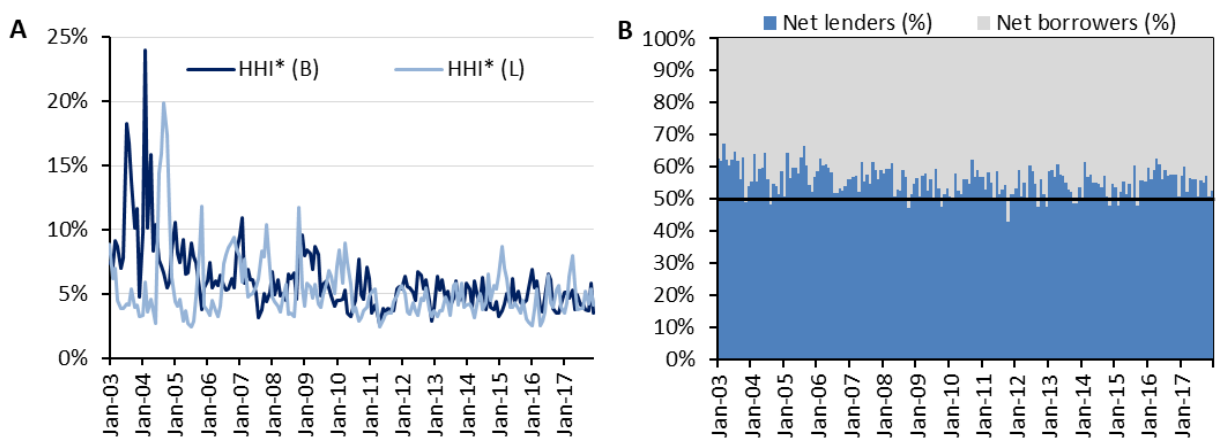

Fig. 7. **(A)** Herfindahl-Hirschman Index (HHI), adjusted by N. **(B)** Distribution of entities according to their monthly net liquidity positions in the market (as percentage of total active nodes in monthly networks).

Another approach to study the network's concentration is based on the analysis of the "net liquidity positions" of the entities in the market. These positions result from netting out all the gross flows traded by each entity during a month. That procedure allows us to elucidate how many banks were net lenders (i.e., provide liquidity to the system) and how many were net borrowers in the market (i.e., absorb liquidity from it). In 91% of the months analyzed there were more net lenders than net borrowers (Fig. 7B), which reflects that the former group was less concentrated than the latter. Nevertheless, the number of net lenders never exceeded 67.5% of total entities, implying that the distribution of liquidity suppliers and demanders was never substantially skewed.

6.3. *Degree distribution*

From a first exploration of the Complementary Cumulative Distribution Functions (CCDFs) of nodes' total degrees, in-degrees and out-degrees (Fig. 8), it can be inferred that they show "heavier" tails than a Poisson. Lognormal distribution appears to be the best fit to most of the observations. However, in the case of out-degrees the evidence also provides some support to the hypothesis of a Power Law behavior in the tail of their distribution. This preliminary approach indicates that there is a large number of banks with few links, coexisting with a small number of highly interconnected entities, which are key to the well-functioning of the network.

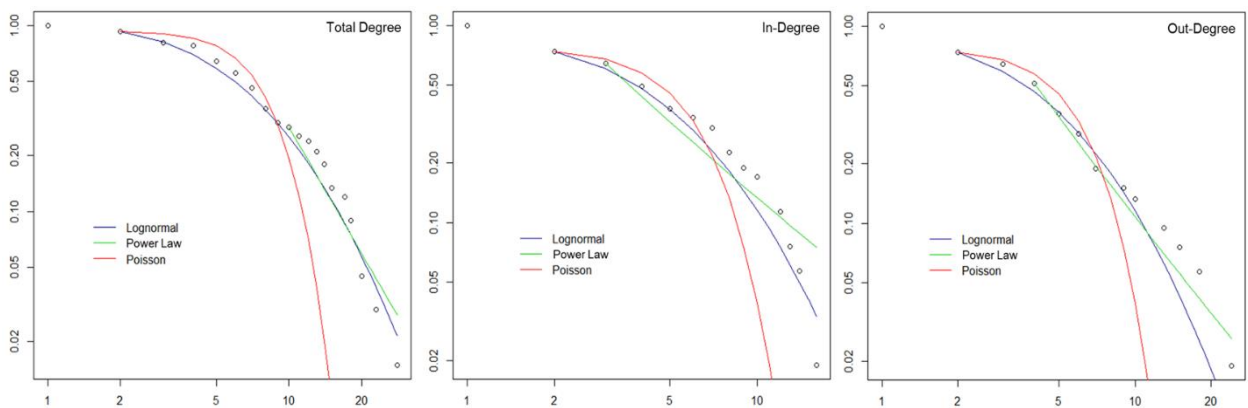

Fig. 8. CCDFs of nodes' degrees, corresponding to the network of December 2016. This month was chosen as a representative example because the topological features of its graph are the most similar to the average of monthly networks over the whole period (67 nodes, 258 edges). Axis are in log-scale.

After applying the methodology introduced by Clauset *et al* [25] to define if the lognormal distribution fits statistically to the monthly empirical networks (and having therefore estimated one p-value per monthly network), it results that in 90% of the cases the null hypothesis of lognormal fit is not rejected for the total degree distribution, considering a type 1 error probability of 10%. Reducing this threshold to 5%, the null hypothesis is not rejected in 95% of the months. Those levels of non-rejection are 88.3% and 92.8%, respectively, for the in-degree distribution, and 91.1% and 95.6% in the case of out-degrees (Table 2). Thus, in almost every month the null hypothesis of lognormal fit is held true, more strongly in the case of the out-degree distribution.

The estimates of $x_{min}$ turned out to be low, which means that the number of observations removed to carry out the estimation of the other parameters is relatively small (compared to the total range of the variable). In fact, the minimum cut-off in the case of total degrees is 4.4 on average (versus maximum of 36 degrees), 3.2 for in-degrees (versus a maximum of 27), and 2.7 for out-degrees (versus a maximum of 23). It is important to highlight again that these are relatively small networks compared to the ones analyzed by other sciences (e.g., biology, physics, linguistics, social networks, etc., where nodes and edges can easily reach thousands or even millions).

Table 2. *Percentage of monthly networks with a degree distribution that does not reject the lognormal hypothesis*

| Stage | Date | Total degree | | | In-degree | | | Out-degree | | |
|---|---|---|---|---|---|---|---|---|---|---|
| | | p-value > 0.1 | p-value > 0.05 | Average Xmin | p-value > 0.1 | p-value > 0.05 | Average Xmin | p-value > 0.1 | p-value > 0.05 | Average Xmin |
| 1 | Jan-03 - Jun-05 | 90.0% | 93.3% | 2.9 | 93.3% | 93.3% | 2.7 | 93.3% | 100.0% | 2.0 |
| 2 | Jul-05 - Dec-07 | 90.0% | 93.3% | 6.3 | 96.7% | 96.7% | 2.7 | 93.3% | 93.3% | 4.3 |
| 3 | Jan-08 - Feb-10 | 92.3% | 92.3% | 3.1 | 88.5% | 88.5% | 2.7 | 88.5% | 88.5% | 1.5 |
| 4 | Mar-10 - Oct-11 | 90.0% | 100.0% | 3.6 | 80.0% | 90.0% | 3.9 | 85.0% | 90.0% | 2.9 |
| 5 | Nov-11 - Nov-15 | 89.8% | 98.0% | 4.5 | 85.7% | 91.8% | 4.0 | 89.8% | 95.9% | 2.9 |
| 6 | Dec-15 - Dec-17 | 88.0% | 92.0% | 5.0 | 84.0% | 88.0% | 3.4 | 96.0% | 100.0% | 2.4 |
| **Jan-03 - Dec-17** | | **90.0%** | **95.0%** | **4.4** | **88.3%** | **92.8%** | **3.2** | **91.1%** | **95.6%** | **2.7** |

The parameter estimates for the fitted distributions can be found in Table 3. When contrasted with other empirical distributions that can be fitted by a Lognormal [28], these parameters are in line with those found in the context of social sciences, economics, linguistics, medicine or geology.

Table 3. *Estimates of Lognormal parameters for the monthly networks' degree distributions*

| Stage | Date | Total degree | | In-degree | | Out-degree | |
|---|---|---|---|---|---|---|---|
| | | Mean | Std. Dev. | Mean | Std. Dev. | Mean | Std. Dev. |
| 1 | Jan-03 - Jun-05 | 1.6 | 0.7 | 1.4 | 0.8 | 0.9 | 0.8 |
| 2 | Jul-05 - Dec-07 | 2.3 | 0.6 | 1.9 | 0.8 | 1.7 | 0.6 |
| 3 | Jan-08 - Feb-10 | 1.7 | 0.6 | 1.4 | 0.6 | 0.6 | 0.7 |
| 4 | Mar-10 - Oct-11 | 1.8 | 0.6 | 1.7 | 0.6 | 1.3 | 0.6 |
| 5 | Nov-11 - Nov-15 | 2.1 | 0.6 | 1.8 | 0.7 | 1.4 | 0.7 |
| 6 | Dec-15 - Dec-17 | 2.0 | 0.6 | 1.8 | 0.6 | 1.3 | 0.7 |
| **Jan-03 - Dec-17** | | **1.9** | **0.6** | **1.7** | **0.7** | **1.3** | **0.7** |

Note: Table 3 shows the average value of the parameter estimates in each Stage, based on the monthly estimates.

To the purpose of comparing the Lognormal goodness-of-fit with that of other distributions usually studied in specialized literature, the log-likelihood derived from the Lognormal hypothesis is contrasted with the resulting log-likelihood of a Poisson and a Power Law, both fitted to the data using an analogous methodology. Applying the same optimal $x_{min}$ that emerges from a Power-Law fit to compute both log-likelihoods, in the 98.3% of cases the log-likelihood derived from the Lognormal fit to the total degree distributions is higher than the one derived from a Power Law. This result also arises in the 99.4% of monthly in-degree distributions and in the 100% of monthly out-degree distributions. From the same procedure, it is concluded that the Lognormal describes better the data than a Poisson in the 96.7% of months in the case of total degree distributions, in the 96.1% of in-degree distributions and in the 97.2% of out-degree distributions.

In sum, the Lognormal distribution, with the parameters shown in Table 3, is the one that best fits to the empirical data, with just a few exceptions. After a detailed examination of these exceptions, no clear regularities emerge that could systematically explain the rejections to the Lognormal hypothesis. They are not concentrated in any of the time stages and those exceptional networks do not show any peculiar topological feature. This heavy-tailed distribution seems to be the most suitable to describe the histogram of total degrees, as well as in- and out-degrees, a phenomenon that has critical implications when assessing the probability of disruptive events and financial fragility, which are key inputs for the design of macroprudential and banking regulation.

## 7. Effects of node centrality on bilateral interest rates

A set of regressions were computed with the aim of measuring the potential effects derived from node centrality on the bilateral interest rates that banks are able to settle in the call market. The dependent variable in this analysis is the interest rate differential between the bilateral rate agreed in each operation and the average market rate in the same day. Formally, the objective is to estimate the impact of node $i$'s centrality (measured by different indicators) on:

$$(6.1) \quad r = \frac{call_{ijt} - call_t}{call_t},$$

where $call_{ijt}$ denotes the interest rate agreed between entity $i$ and entity $j$ on day $t$, and $call_t$ is the average interest rate of all the transactions in the market on day $t$. This specific definition of the dependent variable (as a *percentage* of the market rate, and not in basis points) is useful to avoid biases derived from the substantial volatility experienced by the call rate. On that basis, nine different regressions were carried out using the following generic form:

$$(6.2) \quad r = \alpha + \beta_1 1(centrality\ of\ lender > centrality\ of\ borrower) +$$
$$\beta_2 1(assets\ of\ lender > assets\ of\ borrower) + \gamma_1 D_{lender} + \gamma_2 D_{borrower} + \gamma_3 X$$

$\beta_1$ is the coefficient that quantifies the effect of centrality on $r$. $\beta_2$ measures the impact derived from the fact that the lender has a bigger size than its counterparty (in terms of assets or deposits). These two effects are estimated by the inclusion of binary variables which take the value 1 when the lender is more central that the borrower (in the case of the variable associated to $\beta_1$), or when the former has a bigger balance sheet than the latter (in the case of $\beta_2$), and take the value 0 otherwise. $D_i$ are vectors of dummies included to take into account the type of entities involved in the transaction (in order to control for the differences that arise when SOBs, DPBs, SFBs and/or NBFIs are involved in the transaction). $\gamma_1$ and $\gamma_2$ are the vectors of coefficients associated to this set of dummies. $X$ stands for a vector of control variables related to the conditions of the loan, such as the amount of money (in real terms), maturity in *calendar* days (as all the loans considered have a maturity of 1 *working* day), etc. $\alpha$ is the intercept.

The resulting estimates based on the generic form (6.2) are shown in Table 4. Considering any of the five centrality measures individually (columns 1-5), in all cases their associated coefficients are positive and statistically significant. This implies that a lender that is more central in the network than its counterparty (defining this concept by any of the measures discussed) can negotiate, on average, higher interest rates than the market rate. The most prominent effect is derived from a higher *degree* centrality (column 1), in which case the lender is able to charge interest rates 1.32% above the average call rate. A similar effect, though slightly lower, is caused by strength centrality: it allows the entities to lend at rates 1.11% above the market average (column 5). The other three metrics (closeness, betweenness and eigenvector centrality) appear to be less effective, with an impact of nearly half of the other two measures. In any case, they show positive impacts too, which are statistically and economically significant (columns 2, 3 and 4).

Regarding the type of entity, DPBs are those who lend funds at the highest interest rates, while NBFIs, on average, lend money at the lowest rates. On the borrowers' side, NBFIs tend to obtain the most expensive loans, followed by SOBs in the second place, while SFBs borrow at the most convenient rates. All these coefficients are statistically significant, remain stable across the nine regressions and their sizes reflect the differences between the businesses that each entity runs.

On average, a lender with more assets that its counterparty is able to charge interest rates between 4% and 4.4% higher than the market rate. This coefficient is stable and significant across all specifications, both in statistical and economic terms. If, instead, the size of the entities is measured from the other side of their balance sheets, that means, considering their deposits, the associated coefficient remains practically unchanged (column 8). Overall, these results are in line with those found by similar empirical studies in other countries [3, 21, 23].

Individual liquidity levels also affected bilateral rates. If the lender has higher liquidity levels than the borrower, the agreed rate tends to be 0.2%-0.3% lower than the market rate. Besides, none of the loan characteristics included as controls had significant impacts on $r$ in economic terms.

Table 4. *Dependent variable: interest rate differential between bilateral rates and the average market rate of the day, as a percentage of the market rate ($r$)*

| | Coefficients | | | | | | | | |
|---|---|---|---|---|---|---|---|---|---|
| **Variables** | (1) | (2) | (3) | (4) | (5) | (6) | (7) | (8) | (9) |
| ***Centrality Measures*** | | | | | | | | | |
| *Lender > Borrower, Degree* | 0.0132** | | | | | | 0.0167** | 0.0171** | |
| *Lender > Borrower, Closeness* | | 0.0066** | | | | 0.0052** | 0.0050** | 0.0042** | 0.0049** |
| *Lender > Borrower, Betweenness* | | | 0.0044** | | | -0.0007 | -0.0029** | -0.0003 | 0.0001 |
| *Lender > Borrower, Eigenvector* | | | | 0.0069** | | 0.0052** | -0.0053** | -0.0080** | -0.0011 |
| *Lender > Borrower, Strength* | | | | | 0.0111** | | | | 0.0107** |
| ***Size of financial entity*** | | | | | | | | | |
| *Lender > Borrower, Assets* | 0.0409** | 0.0441** | 0.0441** | 0.0430** | 0.0398** | 0.0431** | 0.0412** | | 0.0411** |
| *Lender > Borrower, Deposits* | | | | | | | | 0.0445** | |
| *Lender > Borrower, Liquidity* | -0.0024** | -0.0027** | -0.0031** | -0.0028** | -0.0028** | -0.0025** | -0.0023** | -0.0004 | -0.0033** |
| ***Type of Lender*** | | | | | | | | | |
| *State-Owned Bank* | -0.0135** | -0.0129** | -0.0127** | -0.0133** | -0.0138** | -0.0132** | -0.0134** | -0.0150** | -0.0127** |
| *Subsidiary of Foreign Bank* | -0.0279** | -0.0267** | -0.0277** | -0.0273** | -0.0281** | -0.0267** | -0.0274** | -0.0269** | -0.0306** |
| *Non-Bank Financial Institution* | -0.0388** | -0.0382** | -0.0390** | -0.0389** | -0.0397** | -0.0381** | -0.0383** | -0.0302** | -0.0378** |
| ***Type of Borrower*** | | | | | | | | | |
| *Non-Bank Financial Institution* | 0.1628** | 0.1671** | 0.1669** | 0.1658** | 0.1644** | 0.1657** | 0.1631** | 0.1680** | 0.1699** |
| *State-Owned Bank* | 0.0530** | 0.0579** | 0.0576** | 0.0564** | 0.0544** | 0.0563** | 0.0534** | 0.0622** | 0.0608** |
| *Domestic Private Bank* | 0.0496** | 0.0518** | 0.0508** | 0.0503** | 0.0477** | 0.0512** | 0.0504** | 0.0518** | 0.0528** |
| ***Loan characteristics*** | | | | | | | | | |
| *Call rate, market average* | -0.0017** | -0.0016** | -0.0016** | -0.0016** | -0.0017** | -0.0016** | -0.0017** | -0.0005** | -0.0005** |
| *Maturity* | 0.000005 | | | | | | 0.000008 | | |
| *Amount* | 0.000005* | | | | | | 0.000005* | 0.0000 | -0.00001** |
| *Days until end of the month* | -0.000002 | | | | | | -0.000003 | -0.0001* | -0.0001* |
| ***Time dummies*** | | | | | | | | | |
| Monthly | Yes | Yes | Yes | Yes | Yes | Yes | Yes | Yes | Yes |
| Annual | No | No | No | No | No | No | No | Yes | Yes |
| **Constant** | -0.0314** | -0.0321** | -0.0300** | -0.0308** | -0.0299** | -0.0328** | -0.0320** | -0.0648** | -0.0525** |
| **Adjusted R²** | 0.1668 | 0.1661 | 0.1659 | 0.1661 | 0.1665 | 0.1663 | 0.1670 | 0.1908 | 0.1875 |

*Significant with a p-value<0.1. **Significant with a p-value<0.01. Each column shows a set of coefficients estimated from a particular specification of (6.2). The $R^2$ was adjusted according to the number of regressors included.

Specifications 6 to 9 regress $r$ on different combinations of centrality measures, to examine their partial effects and potential complementarities. The coefficient associated to betweenness centrality turns out to be unstable and many times non-significant when included jointly with other centrality measures. Something similar occurs in the case of eigenvector centrality. In contrast, closeness centrality's effect remains stable and significant, at around 0.5%, regardless of the combination of variables included. Degree and strength centrality proved to be the most relevant to explain interest rate differentials. In the first case, its associated coefficient even increases when it is included together with other metrics, to a figure near to 1.7%. Strength centrality's effect stands at around 1.1%, also stable and significant even upon changes in the specification.

In conclusion, the evidence seems to support the idea that node centrality represents a relevant factor at the time of negotiating more convenient rates in the call market. The centrality measures with the largest impact are those based on degree and strength, with closeness centrality in the second place. These results highlight the importance of the interconnectedness among financial entities, not only when examining the system from an aggregate perspective but also when approaching the banking business from a micro-level or entrepreneurial point of view.

## 8. Concluding remarks

This paper analyzes for the first time the topological structure of an Argentine interbank market. In general, it exhibits a relatively small size (both in terms of nodes and edges), low density (as is usual in financial networks) and a reciprocity nearly always higher than a comparable random graph. The average distance between banks is lower than 3. The call market is prominently disassortative and displays higher clustering coefficients than a random network of the same size. The 2008 global crisis produced a significant contraction in all these metrics. During the turmoil, banks' reciprocity and clustering levels tended towards those of comparable random graphs, and the network's disassortative mixing decreased in absolute terms in the years after the event.

Certain correspondence was detected between the movements in the size of the network and the country's economic activity. Complementarily, it was found that the number of edges reacted with a positive and high elasticity to changes in the number of nodes. This finding contradicts theoretical models that assume a constant average degree in networks that grow over time.

Regarding the network's degree distribution, the evidence consistently supports the hypothesis that total degrees, in-degrees and out-degrees distributions fit better to a Lognormal than to a Power Law or to a Poisson. The most important implication of this finding is that the empirical degree distributions seem to be heavy-tailed, so they would be incorrectly characterized by a random graph. This means that a narrow group of highly connected banks coexists with a large number of low-degree entities. From a systemic risk perspective, this fact implies that the network tends to be "robust-yet-fragile", in the sense that it is resilient to random failures, but it could be very vulnerable when facing directed attacks to the central nodes. This result is crucial for the design of macroprudential policies, as it highlights the relevance of a rigorous detection of the most central agents, whose failures can potentially disrupt the systemic stability of the market.

Furthermore, it was found that entities with higher centrality in the network tend to agree more convenient bilateral interest rates in their call market operations. Even controlling for the size of the entities, their liquidity levels, type of business and the loan conditions, node centrality explains a non-trivial effect on the capability to lend at a higher interest rate and borrowing at a lower cost.

The topological characterization depicted in this paper posits solid empirical foundations to carry out theoretical exercises and shock simulations, both for the Argentine network and, in general, for analogous markets in similar countries (i.e., middle income countries, with not very developed financial systems). The reported results are valuable to assess to what extent the existing theoretical models about financial networks, contagion, etc., are applicable to the Argentine interbank markets, and therefore to define which are those that best fit to the local dynamics.

The future research agenda is broad. For instance, a deeper analysis of specific stress events of the Argentine history would shed light on potential regularities or stylized facts that might be useful to strengthen systemic stability. More research is needed on other domestic networks, such as the payment system or the cross-holdings of financial assets by different agents. In this line, the study of domestic multilayer networks would be enriching. Complementarily, a dynamic analysis of the banks' trading activity could also provide valuable insights for subsequent theoretical modelling (in the vein of studies like [26]). A thorough understanding of the interdependencies among financial (and non-financial) agents constitutes a crucial task in a world where distances and reaction times are shrinking at a striking pace.

**ACKNOWLEDGEMENTS:** I would like to thank Viktoriya Semeshenko, Marcos Dal Bianco, Daniel Heymann, Martín Zimmermann, Jaromir Kovarik as well as seminar participants at Central Bank of Argentina, University of Buenos Aires, Universidad Nacional del Sur and at the III Interdisciplinary Workshop on Complex Systems (UBA/UDELAR) for helpful comments.

# 9. Appendix: Main topological measures of empirical interbank networks

| Country | Reference | Period | Data type | Frequency | N | M[1] | Density | Reciprocity | Clustering[2] | Assortativity[3] | Average Distance | Degree distribution[4] |
|---|---|---|---|---|---|---|---|---|---|---|---|---|
| Australia | Sokolov *et al.* (2012) | 2007 | Interbank loans | Daily | 55 | 69-83 | 2.6% | - | - | -0.1375 | - | Rejects Power law |
| | | | Other interbank money flows | | 55 | 784-804 | 26.9% | - | - | -0.375 | - | Rejects Power law |
| Austria | Boss *et al.* (2004) | 2000-2003 | Balance sheet exposures | Quarterly | 883 | Max. entropy | - | - | 0.12 ± 0.01 | - | 2.59 ± 0.02 | Power law (α=2.01) |
| Brazil | Cont *et al.* (2013) | 2007-2008 | Balance sheet exposures | Irregular | 592-597 | 1,200 | - | - | 0.2 | Disassortative | 2.35-2.42 | Power law (α=2.54) |
| Canada | Embree & Roberts (2009) | 2004-2008 | Payments | Daily | 14 | - | 69.2% ± 3.3% | 89.3% ± 2.5% | 0.84 ± 0.015 | - | 1.31 ± 0.03 | - |
| Colombia (a) | Cepeda López (2008) | 2006 | Payments | Daily | 126 | 2,245 | 16.4% | 34.2% | 0.61 | - | 2.04 | Power law (Out: α=3.06 / In: α=3.24) |
| Colombia (b) | Machado *et al.* (2010) | 2006 and 2009 | Payments | Daily | 125-137 | 6,843-9,400 | 42.8%-60.6% | - | - | - | 2.04-2.17 | - |
| Denmark | Rørdam & Bech (2009) | 2006 | Interbank money market transactions | Daily | 43.6 ± 4.1 | 75 ± 23 | 11.2% ± 5.8% | 26.2% ± 5.5% | 0.2 ± 0.1 | - | 2.9 ± 0.4 | Exponential |
| | | | Payments | | 89 ± 5.3 | 283 ± 41 | 8.3% ± 0.8% | 22.8% ± 1.8% | 0.5 ± 0.1 | - | 2.5 ± 0.1 | Negative binomial |
| Estonia | Rendón de la Torre *et al.* (2016) | 2014 | Payments | Yearly | 16,613 | 43,375 | 13% | - | 0.183 | -0.18 | 7.1 | Power law (α=2.45) |
| EU | Alves *et al.* (2013) | 2011 | Balance sheet exposures | Yearly | 54 | 1,737 | 60% | 71% | 0.84 | -0.24 | 1.38 | Power law (α=3.5) |
| Germany | Craig & von Peter (2014) | 1999-2012 | Balance sheet exposures | Quarterly | 1,732 ± 85 | 20,081 ± 1,461 | 0.66% | - | - | - | - | Rejects Poisson |
| Hungary | Lublóy (2006) | 2005 | Payments | Monthly | 36 | 774 | 61% | - | - | - | - | - |

| Country | Reference | Period | Data type | Frequency | N | M[1] | Density | Reciprocity | Clustering[2] | Assortativity[3] | Average Distance | Degree distribution[4] |
|---|---|---|---|---|---|---|---|---|---|---|---|---|
| Italy (a) | De Masi et al. (2006) | 1999-2002 | Interbank money market transactions (e-MID) | Daily | 140 | 200 | - | - | $c(k) \propto k^{-0.8}$ | $k_{nn}(k) \propto k^{-0.5}$ | - | Power law ($\alpha$=2.3) |
| Italy (b) | Iori et al. (2008) | 1999-2002 | Interbank money market transactions (e-MID) | Daily | 177-215 | - | - | - | - | Disassortative | - | Not scale-free, but heavier-tailed than a random network |
| Italy (c) | Fricke & Lux (2014) | 1999-2010 | Interbank money market transactions (e-MID) | Quarterly | 120-200 | - | 17%-25% | - | - | Disassortative | - | - |
| Italy (d) | Fricke & Lux (2015) | 1999-2010 | Interbank money market transactions (e-MID) | Daily and Quarterly | - | - | - | - | - | - | - | Negative binomial (daily) Weibull (quarterly) |
| Italy (e) | Kobayashi & Takaguchi (2017) | 2000-2015 | Interbank money market transactions (e-MID) | Daily | 94 | 303 | - | - | - | - | - | - |
| Japan (a) | Inaoka et al. (2004) | 2001-2002 | Payments | Monthly | 354 | 1,727 | 2.76% | - | - | - | - | Power law ($\alpha$=2.1) |
| Japan (b) | Imakubo & Soejima (2010) | 1997 and 2005 | Payments | Monthly | 444 and 354 | 1,383 and 1,709 | 1.4% and 2.7% | - | - | Disassortative | - | Power law ($\alpha$=1.6-3.4) |
| Mexico | Martínez-Jaramillo et al. (2014) | 2005-2010 | Balance sheet exposures | Daily | 27-40 | 280 | 30% | 80% | - | Disassortative | 1.7 | Power law ($\alpha$=3.5) |
| Mexico | | | Payments | | | 471 | 40% | 82% | 0.7-0.85 | Disassortative | 1.5 | Power law |
| Netherlands (a) | Pröpper et al. (2008) | 2005-2006 | Payments | Daily | 129 ± 5 | 1,182 ± 61 | 7% | 63% ± 2% | 0.4 ± 0.02 | Disassortative | 2.0-2.5 | - |
| Netherlands (b) | van Lelyveld & Veld (2014) | 1998-2008 | Balance sheet exposures | Quarterly | 91-102 | ~1,000 | 8% | - | - | - | - | Rejects Poisson and Power law |
| Switzerland | Schumacher (2017) | 2005-2012 | Secured money market transactions | 25-day periods | 161 | - | 10%-20% | 5%-10% | 0.05-0.2 | - | 2-4 | - |
| Switzerland | | | Unsecured money market transactions | | 241 | - | 5% | 20%-30% | 0.1-0.3 | - | 2.6-3.7 | - |

| Country | Reference | Period | Data type | Frequency | N | M[1] | Density | Reciprocity | Clustering[2] | Assortativity[3] | Average Distance | Degree distribution[4] |
|---|---|---|---|---|---|---|---|---|---|---|---|---|
| UK (a) | Becher et al. (2008) | 2003 | Payments | Daily | 337 | 989 | 0.90% | - | 0.23 | - | 2.4 | - |
| UK (b) | Wetherilt et al. (2010)[5] | 2006-2008 | Interbank money market transactions | Daily | 12-13 | - | 42.1%-38.5% | 70.7%-68.9% | - | - | - | - |
| USA (a) | Soramäki et al. (2007) | 2004 | Payments | Daily | 5,086 ± 128 | 76,614 ± 6,151 | 0.3% ± 0.01% | 21.5% ± 0.3% | 0.53 ± 0.01 | -0.31 | 2.6 ± 0.2 | Power law (α=2.11) |
| USA (b) | Bech & Atalay (2010) | 1997-2006 | Interbank money market transactions (federal funds) | Daily | 470 ± 15 | 1,543 ± 72 | 0.70% ± 0.03% | 6.5% ± 0.8% | In: 0.10 Out: 0.28 | -0.06 to -0.28 | In: 2.4 Out: 2.7 | Out: Power law (α=2 ± 0.05) In: Negative binomial |
| Argentina | This paper | 2003-2017 | Interbank money market transactions | Monthly | 65 ± 6 | 237.5 ± 73.1 | 5.5% ± 1.1% | 7.9% ± 3% | 0.19 ± 0.05 | -0.16 ± 0.09 | 2.8 ± 0.5 | Lognormal (μ=1.9 / σ=0.6) |

Notes: 1) maximum entropy refers to a method used to estimate some exposures for which no disaggregated data is available, so the number of links does not emerge directly from the observed information; 2) $c(k)$ accounts for the number of triangles a node of degree $k$ belongs to; 3) $k_{nn}(k)$ is a function that describes the average degree $k_{nn}$ of the neighbors of a node with degree $k$; 4) α refers to the exponent of a Power Law, while μ and σ represent the mean and standard deviation of a Lognormal distribution, respectively; 5) Wetherilt, Zimmerman and Soramäki [46] divided their analysis in two time phases: the first ranges from 18 May 2006 to 8 August 2007, and the second, from 9 August 2007 to 16 December 2008, so their results are reported here separately for each phase. This Table only reports the metrics explicitly mentioned or expressed by the authors of each paper.